**MICHELE TUCCI**

*DEPARTMENT OF PUBLIC ECONOMICS*
*FACULTY OF ECONOMICS*
*UNIVERSITY OF ROME "LA SAPIENZA"*

# EVOLUTIONARY SOCIOECONOMICS: NOTES ON THE COMPUTER SIMULATION ACCORDING TO THE DE FINETTI – SIMON PRINCIPIA

**Abstract**

The present note includes explanatory comments about the synergic interaction within the sphere of the socioeconomic analysis between the two following theoretical frameworks:

1. The Darwinian evolutionary model.
2. The computer simulation in accordance with the principia established by Bruno de Finetti and Herbert Simon.

**Information on the author**: MICHELE TUCCI, Department of Public Economics, Faculty of Economics, University of Rome "La Sapienza".

Address: Via del Castro Laurenziano 9 – 00161 Roma – Italy.

Fax: +39 06 4462040

Email: tucci@dep.eco.uniroma1.it

michele.tucci@uniroma1.it

Web page: http://dep.eco.uniroma1.it/~tucci/

# EVOLUTIONARY SOCIOECONOMICS: NOTES ON THE COMPUTER SIMULATION ACCORDING TO THE DE FINETTI – SIMON PRINCIPIA

by Michele Tucci

**Note for the reader.** The following considerations ought to be regarded as a comment on what the author has treated in the discussion papers Tucci (2002), (2005) and (2006) indicated in the bibliography.

## 1 – The human cultures

Defining the term "human culture" is a difficult task, since such concept refers to an entity implying a vague categorization. Everything that pertains to mankind can be considered as belonging to the human culture. Still, we tend to consider it as a collection of different elements and therefore we suppose that there is more than one culture. Nevertheless, the number is indefinable: there is no way to draw the line dividing one cultural context from the others except by using subjective – and always questionable – criteria. Human societies seem to be structured as a continuum, but then there are chaotic zones where sharp fractures are evident. Once more continuity and discontinuity are just features in the observer's eyes… And the observer is by no means neutral, since every human being must necessarily belong to a specific cultural area. The indetermination principle is at work here: the observer belongs to the phenomenon. The possibility of an objective examination doesn't exist, not even at theoretical level. Probably, the most comprehensive illustration of the nature of the human cultures can be found in the work by Lévi-Strauss "Race et Histoire", where such entity is represented as a huge collection of structures about which it's impossible to define any kind of rank. Each culture is unique and there isn't any chance to create a hierarchy except by relying on a subjective system of values. If we agree with such premises, then it should be obvious that the task of analyzing the evolution of the socioeconomic aspects pertaining to each human environment encounters the same kind of difficulties: each school of thought has its own approach and there is no way to reach a general conformity. It can be evidenced a shared feeling about the fact that every cultural element will eventually change through time and about the importance of reaching an understanding of such phenomena, but then each research tradition goes its own way. A deep scrutiny of these topics is beyond the scope of the present note which is focused on analyzing a specific aspect: the evolution of the technological framework and the correlated transformation of the socioeconomic systems. And we intend to do it by using the Darwinian evolutionary model which is based on the binomial "appearance of the mutation – selection of the fittest". No other types of evolutionary approaches will be taken into consideration.

If we concentrate our attention on the economic aspects of the human cultures, there is one feature that stands out: technological progress is able to break the cultural barriers. It holds the ability to spread throughout dissimilar societies regardless the high level of diversity among each others. Examining what happened in ages that are relatively far from our time, we can find many examples of technological innovations which eventually came to be diffused among almost every human society. Cases of this sort are the great developments of the prehistoric age: the usage of fire, the invention of the wheel – and of weapons like the spear, the bow and arrow or the sword – and the handling of metals. Above all, it stands out the introduction of the agriculture. This last invention is crucial in the human history and it will be examined in a dedicated paragraph. In general term, it ought to be considered one of the greatest socioeconomic transformations, together with the industrial revolution that we are experiencing in the present time and that it's just at the beginning.

The Darwinian evolutionary process – based on the binomial "appearance of the mutation – selection of the fittest" – is able to explain both the development of the technological capabilities and how the economic activities were able to spread the innovations across different cultures. It should be noted that, at least for what history can show us, the innovations were not sufficient to melt the different cultures into a single one. Each human aggregate has been communicating with the others by using the economic sphere as a tool for exchanging knowledge, but each one has kept its own specific features. The existence of the markets, which has been proved even in the very early stages of the human development, must have contributed to such mechanism, but other less peaceful means of communication, like conflicts and invasions, certainly played an important role in spreading the innovations. It should be noted that in spite of the fact that every aspect of the human cultures is supposed to be transformed in the course of time, there isn't yet a satisfactory theory dealing with such phenomena in general terms, while the Darwinian model can explain quite well how the productive structures of the economies evolve. There are instances of societies that created a perfect symbiosis between the economical forms and the other cultural aspects. An illustration can be found in the Egyptian empire which, as Plutarch shows us, is one of the clearest examples of a totally integrated agricultural community. Such situation is not the standard outcome of the Darwinian process of technological development: generally societies tend to be less monolithic and a variety of economic and cultural forms, often conflicting, is more the norm.

Moreover, cultures show aspects which are in the open, clearly visible, and others that operate at unconscious level. Both sides are important in the development of the processes that create innovations: Keynes's "animal spirits" are an example of a category belonging to the second aspect – the obscure one – that plays a crucial role in the evolutionary process. The same can be said about the creative sparkle of the genius that brings a radically innovative idea into the world. This point is relevant to understand how a mutation came to exist and it will be examined in details. But here it should be noted that even if the obscure aspects of the human psyche are relevant both in the changing of the general culture as well as in the evolution of the economic structures, in the second case the Darwinian approach allows us, if not to understand, at least to isolate the role of such feature. In facts, the appearance of at least a mutation is a necessary condition for Darwin's model to operate – since without mutations there cannot be any selection process and therefore the system will be unable to evolve. But the binomial "appearance of the mutation – selection of the fittest" is logically independent from the theory explaining how the mutations are due to appear. Such a degree of intertheoretical autonomy greatly helps us when we proceed to turn general considerations into computer simulations. On the contrary, when dealing with the evolution of the human culture as a whole, we cannot make such a distinction since the material aspects and the contents of the psyche are tangled in an inextricable mixture. Therefore, everything is more difficult…

One interesting feature that can be found among the human societies is the existence of niches, i.e. small human communities that live at a technological level far behind the one prevailing in the rest of the world. Such feature is a strong hint that the evolution of the technological structures of the human aggregates falls into the model illustrated by the Darwinian process. Of course, it's not a proof, since in this kind of problems it's questionable even the usage of the world "proof". The existence of niches shows that the technological evolution is not a natural process, nor it is the surfacing of hidden forces which are supposed to inhabit mankind. It is simply a question of communicating a novelty that has been proved to be the fittest. If a human population lives in isolation, geographically and/or culturally, the only way it can evolve is if an innovative idea would be conceived within the same social group and if the defensive attitude of such a sheltered community would let the innovation emerge. A rather improbable event.

A question that may rise in the mind of the reader is why a theory that was developed for analyzing the evolution of living being proves itself to be so adequate to describe the evolution of technological structures. In the world of the living the Darwinian theory is basically a model of how information is modified and divulged. In facts, when we use the word "mutation" within the realm of biology we mean a change of the information which is contained in each cell of a living being. Moreover, the term "selection" refers to the transmission of information that leads to a fitter individual, while the information associated to unfavorable mutations are discarded. In short, the Darwinian theory based on the binomial "appearance of the mutation – selection of the fittest" operates essentially by mutating and selecting information. Exactly the same mechanism is involved in the Schumpeterian treatment concerning the evolution of the technologies used by the firms in order to produce commodities to be sold on the markets. The Schumpeterian approach is under any respect a Darwinian model. Of course, for the evolutionary process to operate in the world of matter, the information must be incarnated, in the cells of living beings in the first case, in the commodities sold on the markets in the second one. Still, it's always a matter of information been manipulated.

## 2 – The Darwinian model

In the field of socioeconomics the term "evolution" has been employed in so many different connotations that it is virtually impossible to understand its meaning unless we refer to the author that use it. In the present notes we are interested in the Darwinian model of evolution. Darwin's theory is the basic frame of thought for understanding the evolution of the life forms and, since we are on the verge of a new technological revolution based on genetic engineering, obviously it plays a crucial role in the present scientific environment. But there is more to it. Given that our topic is centered on analyzing the evolution of the socioeconomic structures, the focus of our argumentation will be on Darwin thought as an abstract model of "change by chaos", in comparison with the one of "change by law" which is represented by the traditional gravitational framework. Let's proceed to specify some details about those two concepts.

A gravitational model is a dynamical structure showing the same nature of Newton's Universal Law of Gravitation: a mathematical core from which every variable of the system can be calculated. Generally, problems relating to dynamics within a gravitational environment will admit a single and globally stable solution, while exceptions – which may always occur when we move to increase the complexity of the systems under examination – can be analyzed and classified by the usage of the same mathematical apparatus that has been employed to define the model. For example, Thom's catastrophe theory shows that even within well defined gravitational frameworks dynamical events implying multiple solutions can indeed take place. Still, the reason why such occurrences happen is perfectly understandable by examining the nature of the mathematical expressions which define the concerned systems. In other words, we have a vast cosmos with only tiny bits of chaos: a rather reassuring situation.

Darwinian evolution is quite a different case: an ocean of chaos spotted with sparse islands of cosmos. Let's clear one point: if it were possible to find a mathematical law governing the evolution of the living, the academia would scrap Darwin's writings at once: human psyche has a penchant for cosmos, while it fears chaos, even if it's fascinated by it. But such a law doesn't exist and we hold evidence that it may never exists, that it may be out of the realm of possibilities. In the same way, we must admit that a similar state of affairs concerns the evolution of the socioeconomic structures: we don't have any mathematical formula that can be applied to them and chances are that we will never have. In the end, we don't seem to have much of a choice. Since the state of art in socioeconomic studies doesn't provide us with a gravitational scheme that we can use, we are left with the only feasible model of evolution: Darwin's theory.

Let's examine briefly the structure of the Darwinian model and let's sketch it as a general theory, beyond any reference to the nature of the intended applications, by defining the basic categories of the model – in the language of the computer simulation: the agents.

- ▪ The environment: the world where the evolutionary process takes place.

- The individuals: the subjects involved. In Darwin's original theory they are living creatures, but in general terms they may represent elements of a chosen class within the evolutionary world under examination.
- The competitive drive: the forces compelling each individual to struggle for survival.
- The mutation: a modification of the nature of a single individual. The appearance of the mutation and the eventuality that it may spread in the environment are critical features of the theory.
- The selection: only the fittest survives. Probably the most well known aspect of Darwin's thought, but one of the most complex phases of the evolutionary process.

The core of Darwin's approach lays in the binomial "appearance of the mutations – selection of the fittest". Starting with the first one, the appearance of the mutation is the initial condition for an evolutionary process to start. In a perfectly homogeneous environment evolution cannot take place. On the other hand, one point is central: there is no need to have a model explaining why and how mutations appear. Darwin's evolutionary system is perfectly independent from such rationalization. In fact, we may as well suppose that mutations were born from chaos. The only critical condition is that mutations must necessarily exist. Then the selection will start to operate and it's quite evident since the beginning that such process is extremely complex. Firstly, an environment where the selection can take place must be competitive – i.e. there exist forces compelling the agents to satisfy requirements that are critical in order to survive. The next step implies that the mutations which are able to beat the competition continue to exist and are transmitted to the future contexts, while the others disappear. In other words, the individual who is endowed with the winning mutation holds more possibilities than the others to survive and to transmit his favorable innovative feature to his descendants.

At first sight, selection may seem a potentially rational process that may eventually be fully analyzed by the usage of mathematical tools. But this is a deceptive impression which probably stems from our familiarity with the gravitational point of view. In the perspective of a Darwinian evolutionary process, selection is a form of war and we know that war is an event where deterministic and uncertain events conjure to create an utterly chaotic situation. Therefore, this phase of the evolutionary process can be reconstructed only as a case study, since it depends critically on the specific features of the circumstances under examination. Darwin's evolution is a chaotic entity and as such cannot be analyzed by the usage of gravitational tools. At the present state of art, in order to reach a measure of understanding about the course of the events, the best that can be done is to build computer simulations according with the de Finetti – Simon principia.

Finally, it should be noted that with regard to the time scale the Darwinian model shows a fractal nature: it can be used to analyze events that happen rather quickly, as well as processes which lasted for thousands of years. Therefore, we possess an instrument to explore a vast range of socioeconomic phenomena, from the actual ones to those implying historical evolutionary transformations.

### 3 – Scenarios and simulations

Scenarios are stories which hold the power to describe alternative future situations. They are powerful tools which can be employed in forecasting critical events, since their range of application goes way beyond the traditional gravitational prediction methods. Even if such approach can be used in any context, whether it shows a gravitational nature or an evolutionary one, their privileged fields of application belong to the second realm: scenarios are able to anticipate the sudden structural changes that are typical of the chaotic environments. In synthesis, such methodology consists in reconstructing future events to which it is attached a high subjective probability of happening within the analyzed setting. To build useful scenarios we must balance two opposing tendencies: imagination and analysis. The first one is the creative element that is in charge of exploring the potentialities of the future, while the second one must be used to exclude events that are logically impossible. Conceptual instruments to create scenarios are represented by Keynes' "animal spirits" expectations, as well as by de Finetti's "subjective probability" and Simon's "bounded rationality". The common element to the three cited methodologies is represented by subjectivity: the price to be paid to access the power of forecasting.

In fact, one point should be noticed: objective predictions are possible only when we operate in a thoroughly gravitational environment. In any other case, forecasting is a subjective process. A forecasting operation starts always with the creation of scenarios. They are necessary to decide if the phenomena about which we want to predict show enough gravitational properties, so that the traditional methods can be employed with reasonable hopes to reach positive results. On the contrary, if we conclude that the evolutionary elements included in our forecasting problem cannot be neglected, then we are left with only one method: the computer simulation along with the de Finetti – Simon principia. At this point we would need to build some more scenarios in order to define the main features of the simulation: for example the agents that will appear in it, the links among each others and those among them and the environment status variables. Let's examine this matter in details.

A de Finetti – Simon simulation is a structure built "by agents", if we adhere to an expression used in the field of computable socioeconomics, or "by objects" if we employ the computer programming terminology. In any case, the main operation that has to be carried out is the virtual reconstruction – by the usage of a computer programming language – of the critical elements that operate in the environment to be simulated, whether they are individuals, coalitions of individuals, firms, government policies or social tendencies. Such phase is probably the most demanding one within the entire simulation process. Modeling the inner structure of an agent is an act of creation that requires a deep knowledge of the phenomena under examination coupled with the ability of describing complex behavior patterns by using the tools of computer programming. The right attitude toward such goal should be to consider each agent as an intelligent living creature and to investigate its psychological structure with the same care that it's employed by a Jungian analyst in order to scrutinize the archetypes of his patients' psyche. We should adopt the heuristics of an animist shaman: every agent is alive and self-conscious. Therefore, we should analyze it "from the inside", rather than limiting ourselves to take notice of the external reactions. Of course, it's a difficult task, but certainly it can be done. While building the agents' virtual images, a critical point is defining the communication lines linking each agent with every other one. In fact, the map of the interconnections between agents plays a central role in the simulation's dynamics, since it conveys the reaction of the agents in front of the streaming events within the simulated environment. Simon's continuity principle states that the more a simulation is close to the portion of reality we want to study, the more its evolutionary behavior will be similar to what will happen in the real world; nevertheless, since no simulation can be identical to the simulated phenomena, the more we move further on the time scale, the more the evolution of the simulation will differ from the actual events. As a consequence of such principle, limit patterns cannot be analyzed by using Simon's tool. Therefore, asymptotic behavior of a simulation model ought to be regarded as meaningless. The dynamics of the simulation will be limited to a finite number of time intervals which should be able to contain thoroughly the events we want to investigate. If at the end of the selected time period the simulation is unable to provide any answer to the problem we are attempting to solve, then we could try to run the program for an extended number of time intervals. In case we are still unsatisfied, we should reconsider the procedure from the beginning, since chances are that some basic point of the simulation needs radical modifications.

Concerning de Finetti's principium stating the subjective nature of the concept of probability, in the present note we cannot discuss the matter in general terms. What should be pointed out is the inevitability of adopting de Finetti's notion while operating in an evolutionary environment. In fact, in such type of context frequency alone does not suffice to define the probability of a future event, but some additional subjective assumptions are needed. The simplest way to illustrate the point is to build a de Finetti dilemma. Let's suppose that you are a European actuary operating in the middle of the twenties. Your task is to calculate life expectancies in order to set the premiums for life insurances. Indeed a very critical assignment with respect to preserving the financial soundness of the insurance company… Notice that, since you work in a civilized area of the world, you have access to the complete set of data concerning the population records, such as the date of birth and the one of death. Therefore, the dilemma is not a matter of incomplete information. What you could do is using frequencies to calculate probabilities, thus applying a seemingly objective procedure. But then there is a consideration that simply you cannot ignore: only a few years ago there has been a terrible war – needless to say: World War I – that destroyed entire age classes of the population. Therefore, you cannot ignore he fact that your objective frequencies are slanted: they reflect a war context, while on the contrary now you live in peace time. You have a dilemma! There are only two options: either you take the data as they are, and mechanically calculate the required probabilities or you apply a subjective compensation to the data, in order to sterilize the effects of the war as well as you can. The second option is clearly a subjective one. And the first one? If you have a piece of information, and you consciously decide to ignore it even if you know that you are surely making a mistake, isn't that a subjective choice? In fact, there is no way out: in either ways you will be obliged to relish objectivity and to enter the realm of subjective probability. In an evolutionary world coherence requires the usage of subjective probability: in such context, if de Finetti's definition of probability were not to exist, we would not be able to employ the concept of probability at all!

After the virtual structure of each agent is defined and before we let the simulation program run, we will have to define numerically a large number of parameters, most of which involving probabilistic estimates. Such parameters refer to the agent inner features, to the general properties of the simulated environment and to the initially given conditions that must be defined each time we run the program. Whenever it will be necessary, those numerical specifications will be set by using de Finetti's subjective concept of probability.

At this point, it's useful to resume in short the phases of the simulation process.

1. Create a large number of scenarios in order to understand the features of the world we are to explore. In so doing, we delimitate the context of the simulation by defining the elements to be included into it and those which would be redundant or disturbing.
2. Define the inner structure of the agents included in the simulation and draw the map of their interrelations. Of course, the material is to be coded by using a computer programming language.
3. Compile the source code.
4. Eventually by using de Finetti's subjective probability, define all the numerical parameters involved in the process.
5. Run the simulation program as many times as necessary, applying to it all the modifications which should be judged to be useful.
6. Analyze the output and draw the due conclusions.
7. If you are not satisfied, start it all over again.

In the end, it should be noted that creating a successful simulation will require the convergence of several specialized skills. Mainly, three groups of experts are needed: the theorists who can provide the necessary abstractions, the computer programmers who will do the coding and those who have a first hand knowledge of the phenomena that we want to simulate.

**4 – Schumpeter reloaded**

The scope of the present paragraph and of the next one is to define a theoretical environment of evolutionary type in which it would be possible to link in a rigorous way the following three concepts: scientific progress, competition and evolution of the firm productive technology. Of course, the core of the exposition is based on Schumpeter's classical treatment that will be embedded into an explicitly Darwinian framework. In detail, we will sketch the process through which an abstraction in the mind of a creative individual is turned into a more efficient way to produce commodities to be sold on the market. The starting point is an observation that can be applied to a variety of historical circumstances: not always we can detect a meaningful correlation between scientific knowledge, in the form which is traditionally found in universities and advanced research institutions, and the development of the economic system. A recent example can be found in the late Soviet Union, where the level of scientific learning within the academic institutions was rather high, while the performance of the economic structure remained extremely poor. Here we will proceed to support the following thesis: in the above case, the missing variable was the absence of competition within the economic environment. The lack of such feature impeded the process of turning the creative contents within scientists' mind into marketable commodities. In other words, it was lacking the evolutionary phase of transforming human capital into production processes.

Since we are operating in an evolutionary environment, we will proceed to identify within the analyzed framework the binomial condition "appearance of the mutation – selection of the fittest" which constitutes the essence of the Darwinian model. In the economic literature the topic of the selection of the optimal productive structures has been treated along the lines of Schumpeter's analysis, while the crucial problem concerning the way mutations come into existence in the sphere of technologies has rarely been taken into consideration. A feasible approach to our problem can be obtained from a suitable representation of the process that allows mutations to appear. One point needs to be clarified: the appearance of mutation is a necessary condition for an evolutionary process to happen, but it's not a sufficient one. Of course, it's quite obvious that if no mutation appears there cannot be any possibility of evolutionary changes, but the coming into existence of mutations doesn't guarantee that the context under examination will evolve, since there is the possibility that the selection process would discard every innovation, leaving the analyzed environment unchanged. Therefore, in an evolutionary setting we cannot define an algebraic expression establishing a link between the number of mutations and the improvement of the productive structure, since selection may eventually get rid of every mutation that came into existence. Still, we hold the subjective expectation – where the term "subjective" is to be intended as in de Finetti's definition of probability – that the more the mutations, the faster the pace of evolution. Such assumption is based on the fact that we associate a low subjective probability to the case that none among a large number of mutations would be able to pass the screening of the selection phase. We'll come back to this point in a while. Now let's start the reconstruction of Schumpeter's analysis starting from the fundamental category: the technologies that can be employed by the firms for the production of commodities.

Let's take into consideration the set of every productive technology which is available within a given economical environment and let proceed to split the same into two subsets: the one of the "certain" technologies, which are those already in use in order to produce commodities actually sold on the market, and the one of the "uncertain" technologies, which have been experimented in the R & D institutions but have never be used to produce saleable merchandise. Evolution may occur only if at least one of the "uncertain" technologies is put into effective use – thus becoming a "certain" one – and the selection process recognizes it as been more efficient than the previously used one. In order to carry on the following argument, it will be supposed that three levels of assessment will be associated to each "uncertain" technology. Firstly, there is the "ex ante" evaluation which prevails in the R & D world. Then we have the "animal spirits" expectations drafted by the entrepreneurs. It should be noted that the first estimate may substantially differ from the second one. In fact, while the former is essentially derived from the sheltered reality of the R & D labs, the latter must take into account the impact of changes in a complex and partially unpredictable environment: the business world. Finally, if an "uncertain" technology is passed by the selection process, and therefore is turned into a "certain" one, then there is a last evaluation: the "ex post" adding up measuring the actual economic performance of the same.

For the sake of simplicity, let's suppose that each "uncertain" technology bears the same "ex ante" probability to beat the average rate of profit that is associated with the "certain" technologies. As it has already been remarked, such "ex ante" evaluation is related to the R & D world and it may substantially differ from the entrepreneur "animal spirits" estimates. Let's elucidate the case by building a basic example. It will be assumed that within the community of the entrepreneurs each agent will select a single "uncertain" technology that, according to his own subjective point of view, is the most promising in terms of expected gain. Thus, the set of the "uncertain" technologies can be ordered by the number of agents who chose the same element. In other words, to each "uncertain" technology it can be associated a preference index stating the number of entrepreneurs who selected the same. For example, if we suppose that the context under examination includes one hundred entrepreneurs and six "uncertain" technologies, the situation could be as the one represented in the graph (1) – at the end of the present paragraph – where the "uncertain" technologies will be ordered following the degree of preference expressed by the entrepreneurs. It will be presumed that the most preferred "uncertain" technology has been chosen by fifty entrepreneurs, the next one by twenty five and so on. The least favored one has been be selected only by one entrepreneur.

The next and most crucial step implies linking the number of "uncertain" technologies that will be turned into "certain" ones with the level of competition existing in the economy under examination. Therefore, competition will play the role of the control parameter for the coming into existence of mutations. In fact, only when an "uncertain" technology will be transformed into a "certain" one – and thus it will start to be used in the production of commodities to be sold on the market – it will become possible to evaluate the "ex post" rate of return of the same. Generally, the "ex post" measurement of profitability of the new productive structure will differ from both the "ex ante" evaluation of the R & D community and the "animal spirits" expectations of the entrepreneurs.

To complete our reconstruction, we need to introduce another character: the banker. In fact, such agent does not have much to do with banks as they exist in the real world. Rather, he is the virtual personification of the complex structure in charge of setting the level of competition. In a very competitive economy, it can be supposed that every entrepreneur has a direct access to loans and therefore – since the "ex ante" probability of success is exactly the same for every "uncertain" technology – the banker will not discriminate any investment project and he will satisfy each request of financing. Consequently, the whole set of "uncertain" technologies will be turned into new productive structures designed to produce commodities to be sold on the market. Thus, the number of mutations within the examined economic environment will reach the highest possible value.

In less competitive societies the banker will select the "uncertain" technologies to be financed by using a socioeconomic bias. In such a context, the entrepreneurs are unable to act according exclusively to their subjective estimates but they are required to obtain a degree of approval among their "peers". Therefore, the single entrepreneur does not have a direct link with the banker any more. Instead, the latter will set a minimum threshold of consent among entrepreneurs above which it becomes possible to finance an "uncertain" technology. In other words, the banker will finance only those "uncertain" technologies whose preference index is above the minimum threshold. As a consequence of such a choice, a lower number of mutations will appear. In societies that are extremely adverse to competition, the threshold may be set to such a high level that hardly any "uncertain" technology will be turned into a mutation. Considering the exemplification described by the graph (1), we notice that if the criterion for financing is favorable to competition all the six "uncertain" technologies will be turned into "certain" ones. Therefore the number of mutations within the economic system will reach its maximum. On the contrary, if by chance the banker would situate the minimum threshold of consent to twenty, only the "uncertain" technologies I and II would be financed and therefore the number of mutations would be equal to two. Finally, if the banker would set the threshold to sixty, none of the "uncertain" technologies would be transformed into a mutation: every possibility of evolution is excluded.

In order to conclude the present paragraph, we need to stress the importance of a feature which is crucial within the Darwinian model. As already pointed out, the existence of mutations does not guarantee that the system under examination will evolve, since it is possible that the selection process would eliminate every single one, thus maintaining the status quo. Still, we can formulate the following assumption the nature of which is to be considered strictly subjective:

*Assumption (1). In the Darwinian model the pace of evolution of the system is directly correlated with the number of mutations appearing in the environment.*

If we suppose that Assumption (1) is true, then we can connect within our model the pace at which the technological structure of the economy under examination evolves with the level of competition prevailing in the corresponding socioeconomic environment. Consequently, we moved a step further in the solution of the problem formulated at the beginning of the present paragraph. In the next one, we'll go through some more details.

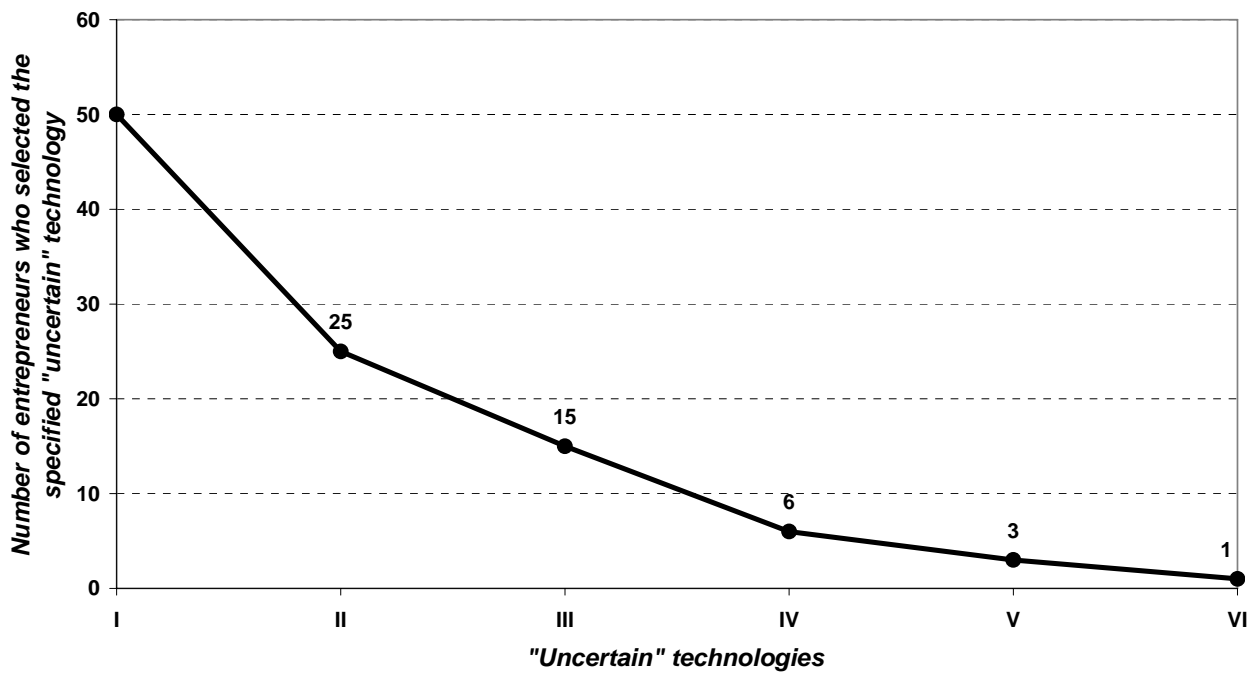

*1 - The "animal spirits" at work*

**5 – Genius**

As already stressed, the Darwinian approach is completely independent from the nature of the process that generates the mutation. In order for an evolutionary process to take place, it's enough that among the mutations that appeared in the environment at least one is able to overcome the selection barrier: if this happens the system evolves. Still, after we sketched how Darwin's approach can be employed to analyze the evolution of the productive structures of a socioeconomic system, we are left with the idea that probably there is something more to be said about the procedure generating the mutations. In fact, if in the biological realm the mutations pertain to the genetic inheritance of each living creature, in the socioeconomic contexts the mutations consist in radically new intuitions. It's an event that starts within the mind of a human being: a sphere about which we hold the presumption to have gained some knowledge. Not enough, probably, but we are not completely unaware of how our psyche works…

The first point to be noticed is that every mutation which eventually will be able to alter the status quo is always conceived by a single individual: a solitary achievement of the mind. The empirical evidence about such statement is overwhelming and probably the supposed exceptions are due to lack of information about how a novelty has been envisaged: it's not always easy to reach the initial source of a critical innovation. On the contrary, what comes after the intuition of the genius is generally a collective work: the R & D phase that produces the "uncertain" technology. Before a radical innovation is turned into a production line manufacturing commodities to be sold on the market, it's necessary to go through a complex sequence which will eventually lead to the evolution of the system. Firstly, there is the phase that starts with an idea in the mind of an individual and ends with an "uncertain" technique which is potentially ready to be used in a standard production line, but it hasn't been employed yet in such role. Generally this process takes place in R & D labs, either at public institutions or at private companies, and often it is financed in ways that do not bear the appearance of capitalistic investments. Vice versa, the procedure of adopting an "uncertain" technique for substituting a production process which is considered obsolete is a typical high-risk capitalistic venture: the investor bets on the eventuality that the selection phase would favor the new arrangement over the old one. Taking into consideration the content of the previous paragraph, the flow chart (2) – at the end of the paragraph – provides a synthesis of the evolutionary path which starts from an innovative idea and leads to major improvements in the productive structure of the economic environment.

The starting point of the evolutionary process – the one that creates the circumstances for the mutations to appear – is obviously represented by the intuition appearing in the mind of a genius who embodies a very peculiar typology of human being. But in fact what is really a genius? For once, the etymology of the word helps us: the origin of the word "genius" is rooted in the old Greek "δαιμων" (*daimon*) which refers to beings in between mortals and gods. In a *daimon* the dual nature, divine and human, leads to a chaotic and unpredictable behavior, both in a positive or negative way, since gods and men cannot mix in an orderly and harmonious manner.

Therefore, the genius is capable of the most astonishing intuitions and, at the same time, he or she may lead a miserable life, materially and/or spiritually. Or simply the genius may behave as such in rare occasions, being for the rest of the time the dullest person of all. Finally, sometimes there may be symptoms of psychosis. The studies about human creativity are a well developed research field and the reader should refer to the specialized literature. Still, one point should be stressed: generally, the sparkle of geniuses lacks in systematization. It may happen only once – or very few times – in the lifespan of the individual and up to now there exist no way to predict such event. Nor to predetermine it. This is the very critical element in order to exert a degree of control over the evolution of the productive structures: if we were able to create an "environment" which would hold the property "to attract" the genius innovative ideas, we would possess a powerful tool to deal with the evolutionary process. Probably, such task would imply finding a way to reward the genius contribution: often the creator gets the cents – if any – while the bureaucrat gets the dollars – a lot. In fact, the other two steps required for the appearance of the mutations are not completely beyond our control. The transition from the innovative intuition to the definition of the "uncertain" technology is a matter of a limited amount of money and the availability of the required R & D facilities. While the last stage – from the "uncertain" technology to the standard production line – depends on whether there is enough venture capital. And whether there are entrepreneurs whose "animal spirits" will induce them to employ the new productive technology on an industrial scale. Then, the phase "selection of the fittest" will occur and there will be winners and losers…

In conclusion, if we are able to understand the Darwinian nature of the transition which leads to the formation of the new productive structures, then we have the chance to create the conditions for controlling the process, and therefore we would be able to deal with the core of the phenomena that influence the evolution of the human societies.

## *2 – The path of evolution*

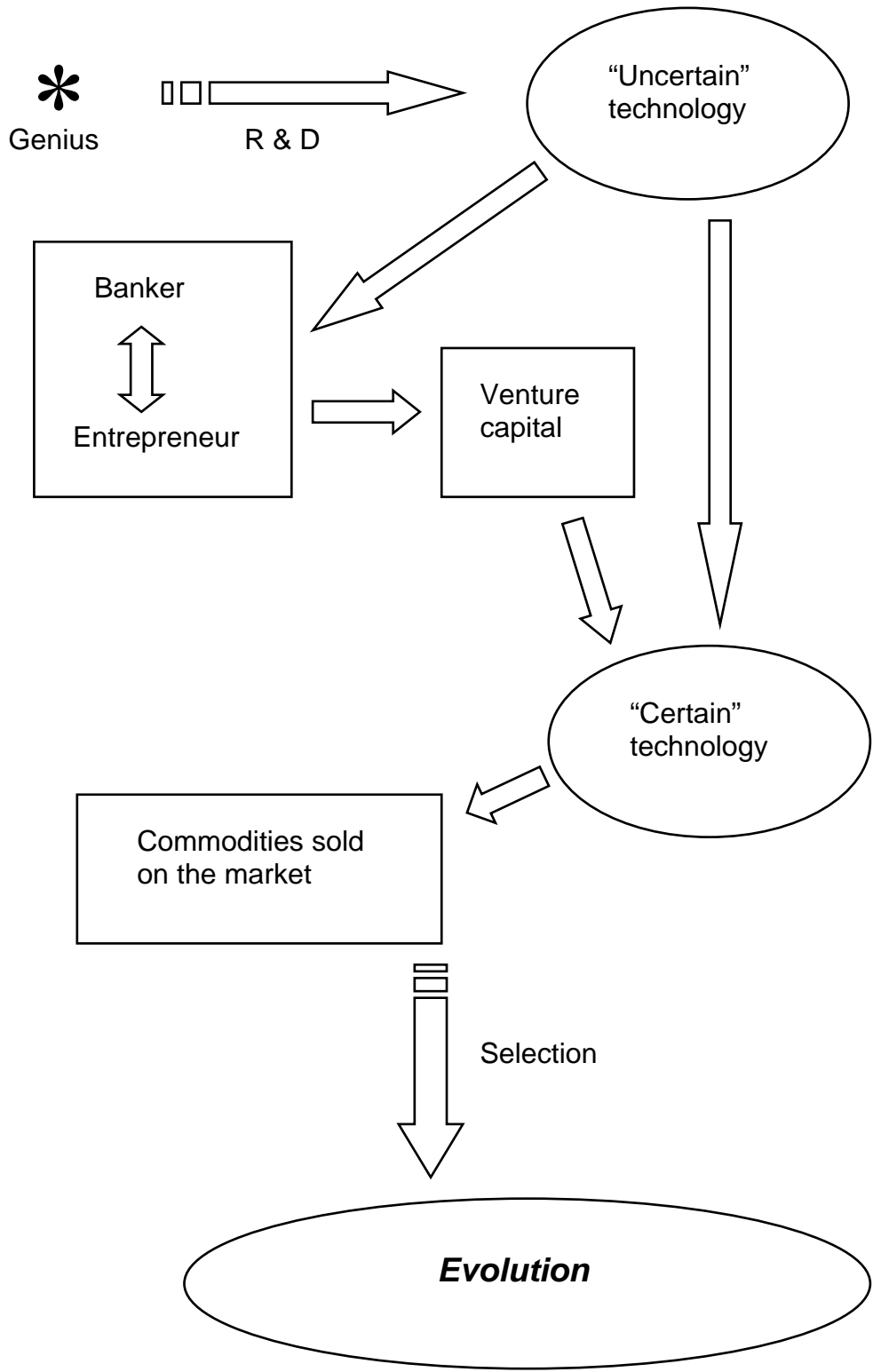

## 6 – Four scenarios

The transition phase between the age of agriculture and the one of industry is not over yet, since a considerable number of human beings still live in what's left of the previous age. Moreover, the latest trends in the more developed areas may suggest that the initial stage of the industrialization – the time of coal and steel – does not represent a faithful image of what is bound to come, since the IT revolution and the globalization of the world economy imply transformations of the socioeconomic structures which were unthinkable just a few decades ago. It is noticeable that the largest part of the socioeconomic analysis carried out by the most prominent intellectuals at the beginning of the industrial revolution appears to be obsolete and it can hardly provide us any valid guideline for understanding the present dynamics.

The industrial age didn't show us its full potential yet. Therefore, we are experiencing a time of quick evolutionary changes which imply the interaction between two lines of metamorphosis: the unfolding capabilities of the new and the painful disappearing of the old. These kinds of evolutionary phenomena show a deeply chaotic nature. As it has already been noticed, the only way to reach a comprehension of such events is by using of the scenario approach at first, then by constructing the suitable computer simulations along with the de Finetti – Simon principia. Thus, it could be interesting to sketch some potential lines of evolution for the times to come: four trajectories going from the worst one to the optimal situation.

***The first scenario: a general collapse of the socioeconomic system*** – This type of events has been so thoroughly explored by sci-fi novels and movies that every reference to it raises a feeling of déjà-vu. Still, such an occurrence cannot be ruled out: it can happen, even if the subjective probability that we attach to it may be rather low. We can choose among a vast collection of catastrophes: giant asteroids hitting the Earth, a dramatic change in the weather conditions, a deadly engineered virus and so on. The fear of a nuclear holocaust has been a constant feature of the post war years. Of course, fiction concentrated on the most picturesque kinds of the destructive happenings, but there are more subtle ones that can equally cause a global crisis. For example, the erosive effects of an ever growing human population could fatally disrupt the process by which the industrial age is finally extending its grasp to the whole humanity. A similar outcome could be caused by the exhaustion of some critical natural resources, prior that the technological progress could find a way to substitute them. An interesting case would be represented by an event subverting the human psyche, rather than the physical conditions of life on the Earth: for example, such an effect could be caused by the take over of a new religious belief that would dramatically alter the actual human behavior. The point with these catastrophic scenarios is that, no matter how many possibilities of annihilation we examine, we don't really believe in any of then. In other words, we know that it can happen, but we assign to these kinds of events a subjective probability near to zero. Still, the socioeconomic system we live in is getting more and more intricate and we know that when the level of complexity raises the possibility of a global crisis increases…

*The second scenario: the "nouveau ancien régime"* – This rather intricate state of affairs is based on the assumption that some social feature of the old agricultural world would mix with the industrial productive structure to bring into existence a mixed structure. In details, it will be supposed that human societies would come to be organized in the rigid pyramidal structure which has been typical of the feudal age, while the hierarchical position of each group would be substantiated, instead that by the possession of land, by the possession of capital. The productive structure would retain its industrial nature, but it would be characterized by an extremely low, if not null, level of technological progress. What kind of evolutionary path would lead to such a situation? The key to understand this scenario rests upon the comprehension of the fundamental ties that link competition among capitalists with technological progress. If we take into consideration the traditional Schumpeterian treatment dealing with the evolution of the productive structures, we can immediately notice that it constitutes an application of the Darwinian evolutionary model, based on the binomial "appearance of the mutation – selection of the fittest". Competition within the industrial system is the key element which allows the transformation of an innovative technology into a standard production line. Therefore, in a world where competition on the markets and in the military sphere would be extremely low, there could not be any significant technological evolution. Such an environment would show some pleasant features: a peaceful humanity prone to solve every problem by negotiations in such a way that the role of armed conflicts would be negligible. It's a world where each person could lead his life without worrying about what would be coming next, because each individual destiny would be socially defined since the time of birth. This last point may not be considered particularly attractive, but after all it's a matter of taste… What should be noticed is that a feudal-like social organization based on a frozen pyramidal society can be compatible with an industrial economy if the level of competition, and therefore of technical progress, is very low: in this sense it can be defined as a "nouveau ancien régime". The negative side of such a scenario may lay in the inefficiency of the economy: the lack of competition does necessarily means that the lower part of the population will be relatively poor. At the end, a question arises: can a socioeconomic structure of this sort be stable? It's difficult to give a general answer to this question, but it's quite probable that a system like the one just being sketched would be affected by a slow but unavoidable decadence, due to the impossibility of serious innovations. And at the end it would be subjected to the curse that Georgescu-Roegen cast in "Energy and Economic Myths": the exhaustion of the Earth capability to sustain human life. There is nothing in this scenario that could prevent such a destiny.

*The third scenario: the everlasting growth* – This setting shows features that are opposite to the one described in the previous point. Competition would be fierce and therefore technological progress would be extremely fast. Human societies would experience a permanent turmoil: wealth and power would pass from an individual to another one without the possibility that stable social classes could be established. Socioeconomic conflicts would be permanent and such a condition would often lead to military confrontations. Individuals would be taken to their physical and psychological limits, causing humanity to progress at the maximum speed, at least if the prevailing chaotic conditions do not lead to one of the catastrophes that has been examined in the first scenario. This contest would show a heroic character and the human colonization of other planets would be taken into serious consideration, since such a constantly fast pace of development would necessarily require so. In this way, Georgescu-Roegen's curse would be exorcised. Since the old agricultural way of life would be abandoned very quickly, the world population should decrease at a relatively fast pace. Still, the negative aspects associated to such dynamics would be more than compensated by the augment of the individual potentialities induced by the accelerated technological developments. A world with less inhabitants, each one well furnished with the most advanced technological supports and ready to compete with the rest of mankind in a win-or-die attitude. In synthesis, it's an exciting way of living, certainly not a relaxed one.

***The fourth scenario: the best of all possible worlds*** – This is the choice of Dr. Pangloss, the pick of the irredeemable optimist. If we examine the last two scenarios, we can notice that, by setting some external criteria, we could try to find the "optimal" values for the parameters that define the scenario. For example, we could proceed to select the level of industrial competition such as the technical progress would be set to the "optimal" value. Then the socioeconomics status variables would reach their "optimal" values: the "optimal" social mobility and the "optimal" distribution of wealth, both aspects balanced by the "optimal" exploitation of the natural resources. Humanity would still need to colonize a new planet someday, but that will happen at the "optimal" time and with the "optimal" methods. Of course, the definition of such "optimal" conditions would necessitate quite a lot of work in order to turn the scenario into a system of equations fit to define the required mathematical maximization problems. Does it sound a bit too familiar? Yes! Indeed it does! The point is that the procedure that has just been outlined basically consists in translating the term "optimal" into the mathematical axioms that would allow us to find the maximizing solution in the form of a unique and globally stable dynamical path. In other words, we are dragged back to the old gravitational way of thinking: looking for an attractor by finding the maximum of a mathematical function. If the formal apparatus has been properly built, we know that probably such a problem can be solved as far as the existence of at least a solution is concerned, while ensuring the uniqueness and the stability of the same is usually an awkward task to be tackled. But the dilemma is: are there some evolutionary agents who are able to force the actual course of the events onto the maximizing path? Or is such an algebraic solution just an abstraction created by a "wishful thinking"? By its own definition, Darwinian evolution cannot be incorporated into a mathematical rule, since the forces that shape evolution appear and disappear during the evolutionary process according to unpredictable factors. None of the two elements that are included in the binomial formula "appearance of the mutation – selection of the fittest" can be calculated by deductive methods. Evolution cannot be analyzed by means of gravitational models.

The scenario of the maximizing path is just the optimist dream. The subjective probability that we attach to it is very near to zero…

## 7 – The population growth: Malthus's curse

The age of agriculture lasted more than ten thousand years. Now it's over. We live in the age of industry that started less than three hundred years ago. In fact, the present time is still a transitional phase, even if probably the change is moving toward completion. Of course, when we use the expression "the age of agriculture" we mean "traditional agriculture", i.e. the type that was practiced before the industrial revolution. Today's agriculture is simply a form of industrial activity and is getting even more so with the deployment of the latest technologies. Traditional agriculture survives in a few areas of the less developed regions and perhaps in some niches of the industrialized word. In any case, it's an economic activity that will become more and more marginal. The main difference between the two forms of agriculture is that the productivity of the first one is limited by natural conditions, while the productivity of the second one can be raised by using industrial know-how: it's a crucial distinction. Traditional agriculture has always been affected by Malthus's curse: explaining this aspect is the scope of the present paragraph. The resulting scenario will be able to deal with the long-term growth trend of the human population and with some cultural aspects of the society organization.

As it's well known, Malthus's proposition states that the growth of population will always outrun the level of the food supply, leading to a decrease in the average supply of food per person. In the agricultural pre-industrial word, Malthus's statement tends to be substantially true. In fact, in those societies the poorest portion of the population has always been on the verge of starvation – or at least chronically troubled by the shortage of food. In every human culture of the old world famine embodied one of the worst demons haunting the collective unconscious. Of course we should consider the criteria of food distribution among the social groups, but it's unlikely that this element could explain the point. Two observations tend to suggest so. Firstly, pre-industrial societies were rather frugal in their lifestyle: it weren't as if the wealthier classes were wasting food. Simply, they had enough of it, while the other section of the population did not. Therefore, the global supply must have been below the requirements. Secondly, it's only very recently – and after that the modern technologies have been applied to the production of food – that one of the most ancient dreams of humanity has become reality: enough food for every human being on earth. In effect, if the distribution were uniform, the actual production of food would result to be slightly more than enough to feed the world. Therefore, it's very unlikely that such achievement could have been reached at the age of traditional agriculture, whose productivity was extremely lower, and whose requirement of manpower was exceedingly higher, than the actual cultivation methods. So, after all, within the set of assumptions that prevailed when he stated his proposition, Malthus was right and Darwin's intuition quite well founded.

As it has been already noticed, the age of traditional agriculture lasted over ten thousand years. It shaped the human destiny up to a point that nearly every civilization existing on earth has been created by the power of the agricultural world. Therefore, we are obligated to formulate a fundamental question: how Malthus's curse affected the history of mankind during such an extended time span? To clarify such issue we should examine a bit more deeply the role of manpower in the agricultural structures. The first issue to be taken into consideration is that traditional agriculture requires quite a lot of work, since the only source of energy comes from human muscles and from some domestic animals. We could imagine that the very idea of labor was born with the invention of agriculture. The switch from the era of the hunting and food gathering societies to the agricultural ones must have been traumatic: a change in the intensity of work and a different concept of time. Hunting and food gathering are activities that require a moderate amount of effort and provide instant satisfaction, while agriculture needs hard work and the result will come in the future…if it will come. After all, agriculture requires faith. We can now sketch the fundamental contradiction – the yin and the yang – that sustained the evolution of those kinds of societies: traditional agriculture requires a lot of manpower, but it cannot feed properly all the people involved in such activity. The reaction is to try and increase the production, but that requires more manpower which will be short of food, and so on. It should be noted that in a traditional environment, due to the virtual absence of technological progress, there existed only two ways to increase the production of food: starting the cultivation of land that has never been cultivated before – which will require even more labor than fertile land – or invading somebody else fields – which means war, a rather man consuming activity. Thus, societies based on traditional agriculture show two apparently conflicting structural features: a lack of food supply and a lack of population. Probably, the second necessity caused the establishment of the patriarchal type of society that characterizes almost every human community based on traditional agriculture. If the target is to increase the population at the maximum possible rate, then women must be relegated to the only role that doesn't admit any substitution: bearing children. Every other social role will be occupied by men. Subjected to these conflicting forces – the reduction of the population due to the lack of food and the maximization of the birth rate by the enforcing of the patriarchal structures – the global human population raised in number very slowly for more than ten thousand years. Stricken by Malthus's curse, the rationale of the age of traditional agriculture has always been to steadily increase the human population. In order to reach this goal, the cultural structures of human societies have been shaped in such way to exert a constant coercion over the mind of the individuals and the communities, enforcing behaviors which were considered in line with such target. The irony is that when this mechanism reached its perfection – i.e. when the human kind had been turned into the flawless device for optimizing the rate of the human reproduction – then the age of traditional agriculture suddenly collapsed and a new way of living was born: the industrial revolution.

For the sake of comparison, it would be interesting to have some information about human communities before the age of agriculture, when the fundamental economic activities were hunting and food collecting. Could it be that the post-modern industrial age bears some resemblance to the era of hunting and food gathering? There is more than one feature creating a link between those two modalities of the human civilization that are so far away in time from each other…Still, satisfying such intellectual curiosity is extremely difficult, since we would like to deal with a contest about which we have very little information. Two different ways for reaching this target have been explored. Firstly, we can try to reconstruct the human environment of that remote age from the few vestiges that we possess: bringing back to life the ur-culture of the Homo sapiens. A fascinating task, but a very hard one to tackle with! Secondly, we can attempt to obtain the same result from the observation of the tiny communities which still today live in a way resembling that of the original hunting and food gathering societies. Both paths are full of difficulties. Nevertheless, there is an amount of agreement among some scholars on the following two points: in the hunting an food gathering communities the level of the population tended to be rather stable and societies were less patriarchal than the agricultural ones: women were present in a larger number of social roles, rather than being restricted to child bearing. For that matter, these beliefs would support the above sketched reconstruction. But let's move to more recent times and let's see how the beginning of the industrial revolution has changed the status quo.

In the last century there happened a remarkable event: in only one hundred years the word population grew by three and half times – year 1900: 1656 mil.; year 2000: 6000 mil.; source: U.S. Census Bureau. At the end of the paragraph graph (3) shows the level of the world population from the year 500 to the year 2000. How and why such occurrence could take place? Following the interpretation that has just been outlined, the answer is rather simple: the industrial revolution started to remove Malthus's curse. Food supplies increased almost everywhere in the world. Therefore, one of the two forces controlling the growth of the population started to disappear, but the other did not. In fact, the patriarchal organization of societies – that relegated women exclusively to the role of child bearers – has lasted a bit longer and only recently it started to collapse in the most advanced industrial countries. Since the only force left operating was the one pushing for the increase of the population, so it did happen: probably we witnessed the fastest augment of the world inhabitants in the whole history of mankind. But now it's fading away. We can notice two symptoms of such inversion of the trend: in the advanced industrial societies we observe both a sharp decrease of the growth rate of the population – sometimes we find a negative rate, if we exclude the contribution of immigration – and a fast change in the society structure, with the vanishing of the patriarchal organization and the access of women to every social position. Moreover, in the developing countries we can often witness the tendency of the social organization to be divided along two structurally different patterns. In the areas where the pressure of modernization is stronger there tend to be present the same features than can be found in the advanced countries: a lower birthrate and a propensity to the relaxation of the patriarchal institutions. On the contrary, the less innovative areas maintain the high birthrate and the patriarchal organization of the past. Since we are considering regions – one in the process of developing and the other holding to the traditional structure – that include population with a homogeneous cultural background, there we have a strong point supporting our opinion.

Now a question rises quite spontaneously. Malthus's curse is something that belongs to the past, even if its effect on the growth rate of the population is still lingering. But is there some feature within the advanced industrialism that is bound to induce a lower level of the population? The answer is probably yes, even if here we cannot examine such assumption in details. But we can provide some hints. It is a fact that at least during the last half century in the leading economies unemployment as been a problem more often than the lack of labor. On the other hand, we can observe that advanced technologies tend to employ less and less labor, while the level of productivity raises at a very fast rate: it's difficult to believe that such type of economic structures will require an increasing amount of labor: eventually the opposite seems more probable. Since the transition from the agricultural system to the industrial one does necessarily imply some chaotic phases, even in the most developed economies we can observe some contexts where the abundance of labor and the consequent low level of wages induce the usage of obsolete productive technologies, which tend to be convenient because of the peculiar structure of the production costs. But the presence of niches is typical of the Darwinian evolutionary processes and such turbulences don't bear a high probability to beat the main trend. Moreover, if we look at the socioeconomic organization of the human societies by employing the Darwinian evolutionary model, there is one symptom that cannot be misunderstood: the fading away of the patriarchal institutions that originally came into existence to guarantee the maximum possible rate of the population growth. In the most advanced industrial countries, more than a fading we can witness a collapse: considering that we are dealing with an institution that lasted thousands of years, the change is unbelievably fast. Such evolutionary transformation could never have happened if it were not a crucial feature of the new era: in the language of Darwin's theory we can state that the alteration embodied in the end of the patriarchal institution has proved itself to be a mutation that was the fittest in the selection process. Therefore, the lowering of the level of the human population is definitely a structural characteristic of the new socioeconomic organization and thus it will spread to the whole world.

At the end of this scenario, we should conclude that we ought to be grateful to Malthus not only for having inspired Darwin in conceiving his evolutionary theory, but also for having clarified a point which could proof itself to be central for understanding the epochal transition typical of the present time.

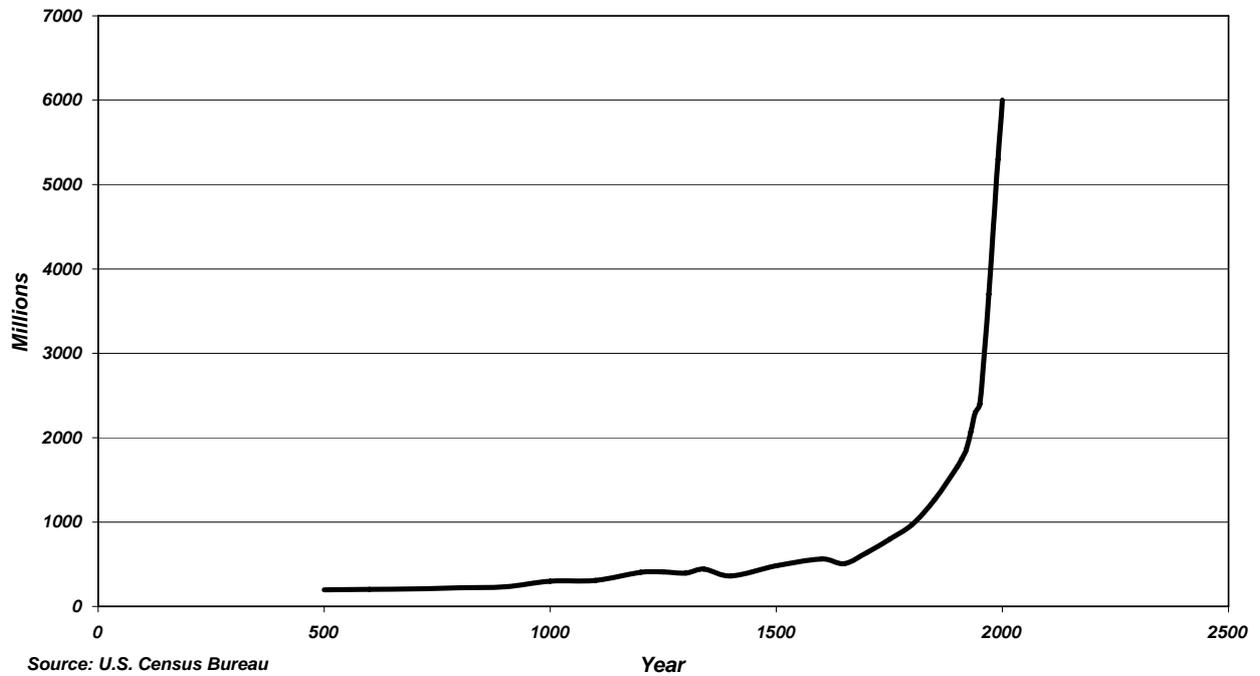

**3 - The world population (years 500-2000)**

Source: U.S. Census Bureau

What is your guess for the next five hundreds years?

**Note**. The bibliography will be expressed in the concise form matching the age of the internet, where information about authors and writings can be obtained instantly.

**References to the author's writings**
The texts of all the articles are available for download at the following address:
http://dep.eco.uniroma1.it/~tucci/
In the same site there can be found the author's CV with the complete bibliography.